\documentclass[a4paper,12pt]{article}
\RequirePackage{ifpdf}
\usepackage{soul}
\usepackage[usenames,dvipsnames]{color}
\usepackage{jheppub}
\usepackage{esvect}
\usepackage{youngtab}
\usepackage{bbold}
\usepackage{float}

\newcommand{\baR}[1]{\overline{#1}}

\newcommand{\Tr}{\text{Tr}}

\newcommand{\ben}{\begin{eqnarray}\displaystyle}
\newcommand{\een}{\end{eqnarray}}

\newcommand{\be}{\begin{equation}}
\newcommand{\ee}{\end{equation}}

\newcommand{\bc}{\begin{center}}
\newcommand{\ec}{\end{center}}

\newcommand{\eesp}{\end{split}}
\newcommand{\bsp}{\begin{split}}

\newcommand{\Rmnum}[1]{\expandafter\@slowromancap\romannumeral #1@}

\newcommand{\g}{\gamma}

\renewcommand{\l}{\lambda}	
\renewcommand{\o}{\omega}	
\newcommand{\q}{\theta}	
	
\renewcommand{\r}{\rho}		

\renewcommand{\t}{\tau}		
\newcommand{\x}{\xi}								

\newcommand{\Z}{\mathcal{Z}}

\newcommand{\cA}{\mathcal{A}}

\newcommand{\cC}{\mathcal{C}}
\newcommand{\cD}{\mathcal{D}}
\newcommand{\cE}{\mathcal{E}}

\newcommand{\cH}{\mathcal{H}}

\newcommand{\cK}{\mathcal{K}}

\newcommand{\cO}{\mathcal{O}}
\newcommand{\cP}{\mathcal{P}}

\newcommand{\cR}{\mathcal{R}}
\newcommand{\cS}{\mathcal{S}}
\newcommand{\cT}{\mathcal{T}}

\newcommand{\cZ}{\mathcal{Z}}

\newcommand{\ra}{\rightarrow}

\newcommand{\lB}{\left [}
\newcommand{\rB}{\right ]}
\newcommand{\lb}{\left (}
\newcommand{\rb}{\right )}

\newcommand{\for}{\text{for}}
\newcommand{\where}{\text{where}}
\newcommand{\with}{\text{with}}
\newcommand{\tand}{\text{and}}

\newcommand{\bensp}{\begin{eqnarray}\begin{split}}
\newcommand{\eensp}{\end{eqnarray}\end{split}}

\newcommand{\bnm}{\begin{matrix}}
\newcommand{\enm}{\end{matrix}}

\def\XXint#1#2#3{{\setbox0=\hbox{$#1{#2#3}{\int}$ }
\vcenter{\hbox{$#2#3$ }}\kern-.6\wd0}}

\newcommand{\ket}[1]{|#1\big>}
\newcommand{\bra}[1]{\big<#1|}

\newcommand{\braket}[2]{\big<#1|#2\big>}

\newcommand{\dow}{\partial}

\begin{document}
	
	\title{From 2d Droplets to 2d Yang-Mills}
	\author{Arghya Chattopadhyay$^a$,}
	\author{Suvankar Dutta$^{b}$,}
	\author{Debangshu Mukherjee$^{b,c}$,}
	\author{Neetu$^b$}

	\affiliation{$^a$Institute of Mathematical Sciences, Homi Bhaba National Institute (HBNI) \\IV Cross Road, Taramani, Chennai 600113, Tamil Nadu, India}
	\affiliation{$^b$Indian Institute of Science Education and Research Bhopal\\
	Bhopal Bypass, Bhopal 462066, India } 
	\affiliation{$^c$Indian Institute of Science Education and Research Thiruvananthapuram\\
	Vithura 695551, Kerala, India} 
	\emailAdd{arghyac@imsc.res.in}
	\emailAdd{suvankar@iiserb.ac.in}
	\emailAdd{debangshu@iisertvm.ac.in}
	\emailAdd{neetuj@iiserb.ac.in}

\abstract{We establish a connection between time evolution of free Fermi droplets and partition function of \emph{generalised} \emph{q}-deformed Yang-Mills theories on Riemann surfaces. Classical phases of $(0+1)$ dimensional unitary matrix models can be characterised by free Fermi droplets in two dimensions. We quantise these droplets and find that the modes satisfy an abelian Kac-Moody algebra. The Hilbert spaces $\cH_+$ and $\cH_-$ associated with the upper and lower free Fermi surfaces of a droplet admit a Young diagram basis in which the phase space Hamiltonian is diagonal with eigenvalue, in the large $N$ limit, equal to the quadratic Casimir of $u(N)$. We establish an exact mapping between states in $\cH_\pm$ and geometries of droplets. In particular, coherent states in $\cH_\pm$ correspond to classical deformation of upper and lower Fermi surfaces. We prove that correlation between two coherent states in $\cH_\pm$ is equal to the chiral and anti-chiral partition function of $2d$ Yang-Mills theory on a cylinder. Using the fact that the full Hilbert space $\cH_+ \otimes \cH_-$ admits a \emph{composite} basis, we show that correlation between two classical droplet geometries is equal to the full $U(N)$ Yang-Mills partition function on cylinder. We further establish a connection between higher point correlators in $\cH_\pm$ and higher point correlators in $2d$ Yang-Mills on Riemann surface. There are special states in $\cH_\pm$ whose transition amplitudes are equal to the partition function of $2d$ \emph{q}-deformed Yang-Mills and in general character expansion of Villain action. We emphasise that the \emph{q}-deformation in the Yang-Mills side is related to special deformation of droplet geometries without deforming the gauge group associated with the matrix model.
}

\maketitle


\section{Introduction and summary}
\label{sec:intro}

The theory of random matrix integrals has achieved so many accolades that a very few of the other \emph{toy models} can ever reach. Random matrix integrals (in short matrix models) have impacted both the realms of mathematics and physics almost equally. Although, matrix models have not been very helpful in the context of critical string theory or superstring theories but one of the most significant model, the \emph{$\mathit{c=1}$ matrix model} alias \emph{matrix quantum mechanics} (MQM) did serve as an indispensable tool to study lower dimensional bosonic string theories. Conventionally, the name ``$c=1$ matrix model'' is derived from the fact that the double scaling limit of this one dimensional Hermitian matrix model represents a two-dimensional string theory whose target space interpretation is that of a Liouville theory coupled with $c=1$ matter. Early evidences of such connection were explored in \cite{POLYAKOV1981207,POLYAKOV1981211} and made robust in the early 90's by Gross and Klebanov \cite{GROSS1991459}. For a more comprehensive overview, the reader may refer to \cite{Klebanov:1991qa,Mukhi:2003sz}. Following the beautiful works laid down by \cite{Menotti:1981ry,Gross:1993hu}\footnote{See also \cite{Boulatov:1992pk,Baez:1994gk,Cordes:1994sd,Horava:1993aq,Gross:1992tu,Gross:1993yt} for more works on $2d$ Yang-Mills and string theory.}, it was shown in \cite{Minahan:1993np} that the same model can also describe a two dimensional Yang-Mills theory with compactified spatial dimension. More interestingly \cite{Minahan:1993np} proved that the same $2d$ Yang-Mills partition function can be rewritten as an one dimensional unitary matrix model or a \emph{unitary matrix quantum mechanics} (UMQM) for both $U(N)$ or $SU(N)$ gauge groups\footnote{See \cite{Ramgoolam:1993hh} for connection between $O(N)$ or $Sp(N)$ $2d$ Yang-Mills theory and string theory.}. In \cite{Caselle:1993gc} this observation was extended  to include $2d$ Yang-Mills defined on both cylinder and torus. One of the main aims of this paper is to further investigate the correspondence between UMQM and $2d$ Yang-Mills defined on higher genus Riemann surfaces.

$2d$ Yang-Mills partition function with gauge group $G$ on a Riemann surface $\Sigma_g$ with genus $g$ can be written as \cite{Cordes:1994fc}
\begin{equation}\label{eq:QCDpart}
    \cZ_{\Sigma_g}=\sum_{\cR} \dim \cR^{2-2g}e^{-{\beta\over 2}C_2(\cR)},
\end{equation}
where the sum is over all possible representations $\cR$ of $G$. $\dim \cR$ is the dimension of $\cR$. $C_2(\cR)$ is the quadratic Casimir of $\cR$ and $\beta$ is a theory dependent constant. Due to lack of propagating degrees of freedom in $2d$ Yang-Mills, the interesting objects to study in this theory are the correlators of Wilson loops $W(U)=\Tr U $ where $U= \text{P}\,e^{\oint_\gamma A}$ around different edges (or boundaries) of $\Sigma_g$. Following \cite{blauthomp,Witten:1991we} these correlators are given by
\begin{equation}\label{eq:heatkernel}
    \cK_{g,n}\equiv \langle W(U_1)\cdots W(U_n)\rangle =\sum_\cR \dim \cR^{2-2g-n} \chi_\cR(U_1)\cdots \chi_\cR(U_n)e^{-{\beta\over 2}C_2(\cR)},
\end{equation}
where $\chi_R(U_n)$ is the character of the representation $\cR$ evaluated on the holonomy $U_n$. One can then simply recover (\ref{eq:QCDpart}) from (\ref{eq:heatkernel}) by setting $n$ to zero.

Starting with the partition function (\ref{eq:QCDpart}) and considering the gauge group $G$ to be $U(N)$ or $SU(N)$, \cite{Minahan:1993np} constructed the Hamiltonian by showing that the states in the theory consist of interacting strings that wind around the circle. They also observed that the Hamiltonian is equivalent to Das-Jevicki Hamiltonian \cite{das-jevicki,sakita} of the $c=1$ matrix model with the condition that the spatial direction is compactified. This provides an equivalence between $2d$ Yang-Mills and UMQM. In an alternate approach, \cite{Caselle:1993gc} showed that $2d$ Yang-Mills partition function on a cylinder or torus is exactly same as one dimensional Kazakov and Migdal matrix model \cite{Kazakov:1992ym} with eigenvalues distributed on a circle and can be interpreted as UMQM. Another way to provide a free-fermionic description of $2d$ Yang-Mills is using quantum mechanics on group manifold $U(N)$ \cite{Douglas:1993wy}. This is arguably the most useful approach to the problem since UMQM are nothing but a system of $N$ free fermions, and therefore the classical limit ($N\rightarrow \infty$) can be best described in terms of phase space variables. 

Contrary to the usual, in this paper we take a \emph{bottom-up} approach to show the equivalence between unitary matrix quantum mechanics and $2d$ Yang-Mills theory and its variants via quantisation of phase space droplet. We explicitly construct the partition function of $2d$ Yang-Mills theory on a generic Riemann surface with gauge group $U(N)$ or \emph{q}-deformed $U(N)$ from the evolution of free Fermi droplets in one dimensional unitary matrix models. To be precise, we quantise the classical droplet in the matrix model and construct the corresponding Hilbert space. The Hilbert spaces $\cH_+$ and $\cH_-$ associated with the upper and lower free Fermi surfaces of a droplet admit a Young diagram basis in which the phase space Hamiltonian is diagonal with eigenvalue, in the large $N$ limit, equal to the quadratic Casimir of $u(N)$. We establish an exact mapping between the states in $\cH_\pm$ and the geometries of upper and lower Fermi surfaces. In particular, coherent states in $\cH_\pm$ correspond to classical deformation of upper and lower Fermi surfaces. We then prove that correlation between two coherent states in $\cH_\pm$ is equal to the chiral and anti-chiral partition function of $2d$ Yang-Mills theory on a cylinder, establishing the fact that factorisation of $2d$ Yang-Mills in chiral and anti-chiral sectors is equivalent to independent evolution of upper and lower Fermi surfaces. More generically we prove that chiral $2d$ Yang-Mills partition function on a generic Riemann surface with $n$ punctures (\ref{eq:heatkernel}) have a rather simple interpretation in terms of correlations between coherent states in $\cH_\pm$. Using the fact that the full Hilbert space $\cH_+ \otimes \cH_-$ admits a \emph{composite} basis, we show that correlation between two classical geometries is equal to the full $U(N)$ Yang-Mills partition function on cylinder. There exists a special class of coherent states in the Hilbert space - we call them \emph{q}-deformed coherent states. \emph{q}-deformed coherent states are mapped to a particular geometries of Fermi surfaces - \emph{box} shaped surface. Correlation between these \emph{q}-deformed coherent states is related to the character expansion of Villain action studied in \cite{Onofri:1981qk}. Since the character expansion of Villain action can be related to \emph{q}-deformed Yang-Mills theory \cite{Romo_2012}, our analysis gives a dictionary between the correlators in UMQM Hilbert space and the heat kernel of \emph{q}-deformed Yang-Mills. We emphasise that the \emph{q}-deformation in the Yang-Mills side is related to special deformation of geometries without deforming the gauge group associated with the matrix model. Therefore the droplet picture of unitary matrix model is much more general and vivid. In some sense the geometry of droplet unifies different versions of $2d$ Yang-Mills theories. A droplet contains more information than it is expected.

We now elaborate our results in detail.  The camaraderie between representation theory and quantum field theory of free fermions, exploited earlier by \cite{Minahan:1993np,Douglas:1993wy} is again at the core of our current approach. We start with a generic $(0+1)$ dimensional unitary matrix model. The dynamics of eigenvalues can be described in terms of a collective field $\rho(t,\theta)$ and its conjugate momentum $\pi(t,\theta)$ \cite{sakita}. Classical dynamics of collective fields also admits an equivalent description in terms of evolution of free Fermi droplet \cite{polchinski}. This two pictures are related by the simple fact that solving Hamilton's equations for collective field and its conjugate momentum is equivalent to finding upper and lower Fermi surfaces of a free Fermi droplet (denoted by $p_+(t,\theta)$ and $p_-(t,\theta)$ respectively in this paper). The advantage of the second picture is that the equations of motion for $p_+(t,\theta)$ and $p_-(t,\theta)$ are decoupled and hence their evolution, except for the fact that $p_+(t,\theta)-p_-(t,\theta) \geq 0$ always as the difference is equal to eigenvalue density up to a factor of $2\pi$. The collective field theory Hamiltonian while written in terms of $p_\pm(t,\theta)$ becomes diagonal (separable). Thus finding eigenvalue configuration is equivalent to find the shape/geometry of the free Fermi droplet. Unitary matrix quantum mechanics with zero potential admits a classical solution $p_\pm =\pm \frac12$. This solution corresponds to a uniform droplet and a constant eigenvalue configuration. In this paper we quantise this classical solution/droplet. However quantisation of other classical solutions can also be done in a similar way. We list our main observations sequentially.
\begin{itemize}
    \item {\bf Quantisation of droplets :} We quantise the classical droplet by imposing equal time commutation relations on $p_\pm(\theta,t)$. Expanding $p_\pm(\theta,t)$ in Fourier modes, we find that the modes in $+$ and $-$ sectors individually satisfy the abelian Kac-Moody algebra and the modes in two different sectors commute. Similar quantisation was studied by \cite{Jevicki:1996fd} in the context of Hermitian matrix quantum mechanics. Writing the Hamiltonian in terms of these modes we find that the Hamiltonian contains a usual quadratic piece (free part) as well as a cubic piece (interaction terms). This Hamiltonian is exactly the same as the one obtained by \cite{Minahan:1993np,Gross:1992tu, Douglas:1993wy}, with the cubic term being similar to the string splitting-joining interaction. 
    \item {\bf The Hilbert space $\cH$ :} We construct the Hilbert space associated with the quantised droplets. Demanding area preserving time evolution of the droplet, we find that the zero modes of $+$ and $-$ sectors are equal up to a sign. This constraint along with the fact that the zero modes commute with the Hamiltonian suggests that the Hilbert space can be constructed upon a one parameter family of ground state $\ket s$. We take the parameter $s$ to be integer, such the the phase space momentum is quantised. A generic state in the Hilbert space is given by action of creation operators associated with $p_\pm$ on the ground state $\ket s$.
    \item Since $+$ and $-$ sectors are decoupled, the excitation in $+$ sector are isomorphic to those in the $-$ sector. The full Hamiltonian is also given by a direct sum of $H_+$ and $H_-$ : $H=H_+ + H_-$. As a result, the total Hilbert space is a direct product of Hilbert spaces for $+$ and $-$ sectors : $\cH = \cH_+\otimes \cH_-$. Evolution of states in $\cH_+$ and $\cH_-$ are independent of each other.
    \item {\bf Mapping between $\cH$ and droplet :} We then define a mapping between a state $\ket{\psi} \in \cH_\pm$ and geometries of upper and lower Fermi surfaces respectively : $\ket{\psi} \ra \{\bra{\psi}p_\pm(\theta)\ket{\psi}\}$. The mapping is one-to-one as long as $p_\pm(\theta)$ is a single valued function of $\theta$. The expectation value of $p_\pm$ in the ground state is $\bra{s}p_\pm \ket{s} = \pm \frac12 + \frac{s}{N}$ with zero dispersion. Therefore the ground state $\ket s$ corresponds to an overall shift of $p_\pm$ by an amount $s/N$ over the classical shape. Since the eigenvalue density is proportional to the difference between $p_+$ and $p_-$, such constant shift in upper and lower Fermi surfaces does not change the eigenvalue distribution. Expectation value of $p_\pm$ in a generic normalised excited  state is same as the expectation value in the ground state. However the quantum dispersion $\Delta p_\pm$ in a generic excited state is not zero and goes like $1/N$. Therefore, these states are quantum excitations over the classical shape. In \cite{Minahan:1993np} such excitations in $2d$ Yang-Mills were identified with the left and right winding of strings around the circle. These excited states form a basis in the Hilbert space for a given ground state $\ket s$. Since $+$ and $-$ sectors are decoupled and isomorphic, we consider excitations in $+$ sector only before combining the two sectors.
    \item {\bf Coherent states in $\cH_+$ :} We also construct \emph{coherent states} in the Hilbert space $\cH_+$. Expectation value of $p_+$ in a coherent state is non-zero and gives a finite deformation of $p_+$ over the ground state. Quantum dispersion of $p_+$ in a coherent state is zero. Hence we call such states \emph{classical}. A classical state preserves the quadratic nature of droplets : for a given $\theta$, $p_+$ is unique.
    \item {\bf Diagonalisation of the Hamiltonian :} It turns out that the interaction piece in the Hamiltonian (of either sectors) is not diagonal in the above basis. We define a new basis in terms of representations of permutation group, in which the Hamiltonian is diagonal \cite{Douglas:1993wy}. In representation basis $\cH_\pm$ are spanned by representations built out of fundamental or anti-fundamental respectively. In $N\ra \infty$ limit the eigenvalue of the Hamiltonian $H_\pm$ in representation basis is equal to the second Casimir of $u(N)$ representation. The Hamiltonian $H_\pm$ is same as that of a $U(N)$ $2d$ Yang-Mills theory in chiral (anti-chiral) sector. 
    
    \item {\bf Evolution of coherent states in $\cH_+$ and the cylinder amplitude :} We next study the evolution of classical states in the Hilbert space. We show that the transition probability of an initial coherent state to a final coherent state in time $T$ is same as the chiral partition function of a $U(N)$ $2d$ Yang-Mills theory on a cylinder with two specified holonomies. In the droplet picture such a propagator gives the evolution of upper Fermi surface from one classical shape to the other. The initial and the final shapes are therefore mapped to two holonomies in the $2d$ Yang-Mills side.  
    \item {\bf Disc, sphere and torus :} There exists a special coherent state which corresponds to $p_+(\theta) \sim \delta(\theta)$. The transition amplitude of a generic coherent state to this special state is mapped to chiral \emph{disc} partition function of $2d$ Yang-Mills theory. If both the initial and final shapes are delta functions, then such propagators are mapped to sphere partition function of the $2d$ Yang-Mills in the chiral sector. Defining appropriate surgery in the space of coherent states we also generate the torus partition function. 
    \item {\bf Higher point correlators :} We define a unique operator associated with a coherent state. $n$-point correlations of these operators in a mixed ensemble (weight factor depends on an integer $g\geq 0$) in the Hilbert space $\cH_+$ can be related to a generic Yang-Mills amplitude (\ref{eq:heatkernel}) in a chiral sector. For example transition amplitude between two coherent states, discussed above is same as two point correlator in a mixed ensemble with $g=0$. 
    \item {\bf \emph{q}-deformed theory :} The correspondence between the correlators of coherent states and heat kernel of $2d$ Yang-Mills can be naturally extended to $q$-deformed Yang-Mills theory. We define a new class of coherent states as a function of $q$. Projection of such states along representation basis gives $q$-deformed dimension of the corresponding representations \cite{Douglas:1993wy}. Considering the deformation parameter $q=e^{i g_s}$ and taking the double scaling limit $g_s\ra 0, \ N\ra \infty$ keeping $g_s N=\lambda$ fixed, the expectation value of $p_+$ operator in such states is given by a box function symmetrically distributed about $\theta =0$ with height $1/\lambda$ and width $\lambda$. With the aid of these special class of coherent states we are able to reproduce the sphere and disc partition function of \emph{q}-deformed Yang-Mills. The sphere partition function corresponds to evolution of a \emph{q}-deformed coherent state to itself. Evolution of a \emph{q}-deformed coherent state to generic coherent state maps to a disc partition function. 
    \item {\bf Connection with Villain action :} Going a step further, we also compute the correlation between two different \emph{q}-deformed coherent states. In the droplet picture such a correlation corresponds to evolution of a box distribution to another box distribution keeping the area preserved. These correlators do not have any direct consequence in $q$-deformed theories. However, we show that they appear in the context of character expansion of Villain action \cite{Onofri:1981qk}. Integrating two different $U(N)$ Villain actions (with two different parameters) over $U(N)$ group manifold one obtains such amplitudes. 
    \item {\bf Joining two chiral sectors :} In order to recover the full $2d$ Yang-Mills partition function, one has to consider the contribution of propagators coming from states in both the Hilbert spaces $\cH_+$ and $\cH_-$. Tools for factorizing Yang-Mills theories into chiral and anti-chiral sectors have been developed using the notion of composite representations.  We show that the full Hilbert space $\cH_+\otimes \cH_-$ admits  a composite representation basis.   Expressing  the  evolution amplitudes in this basis we recover the full partition of $U(N)$ Yang-Mills theory in $2d$.
\end{itemize}

We have structured the paper as follows. In section \ref{sec:umqm}, we briefly discuss the construction of classical phase space in matrix quantum mechanics. Quantisation of classical phase space is given in section \ref{sec:quantisation}. The connection between droplet evolution and partition functions of (\emph{q}-deformed) $2d$ Yang Mills theory is discussed in section \ref{sec:correlation}. Further in appendix \ref{app:eigenphase}, we provide the eigenvalue analysis for the phase space Hamiltonian. A brief review of composite representation and some comments on irreducible representations of $su(N)$ and $u(N)$ appears in appendix \ref{app:repgyan}. In appendix \ref{app:higherpoint}, we elaborate on twisted surgery and explicitly write down an expression for 3-point function (however the process can in principle to generalized to obtain general $n$-point functions). Finally, in appendix \ref{app:villain} we discuss about the Villain action and its character expansion.


\section{Unitary matrix model in $(0+1)$ dimension} \label{sec:umqm}

Partition function for a unitary matrix model in $(0+1)$ dimension is given by
\ben\label{eq:mmqmech}
\Z_t = \int [DU] \exp\lB \int dt \lb \Tr\dot U^2 
+ W(U)\rb \rB.
\een
Following the beautiful work by Jevicki and Sakita \cite{sakita, jevicki} one can describe the matrix model (\ref{eq:mmqmech}) in terms of a real collective bosonic field $\rho(t,\theta)$ (the eigenvalue density) and its conjugate momentum $\pi(t,\q)$. The corresponding Hamiltonian is given by,
\be\label{eq:HamB}
H_B = \int d\q \lb \frac12 {\dow \pi(t,\q) \over \dow\q} \r(t,\q)
{\dow \pi(t,\q) \over \dow\q} + \frac{\pi^2 \r^3(t,\q)}{6} + W(\q)
\r(t,\q) \rb
\ee
where $\int d\theta W(\theta) \rho(t,\theta)= W(U(t))$. The Hamilton's equations for $\rho(t,\theta)$ and $\pi(t,\theta)$ are given by,
\ben
\begin{split}\label{eq:colleom}
\dow_t \r(t,\q)+\dow_\q \lb \r(t,\q) v(t,\q)\rb &=0\\
\dow_t v(t,\q) + \frac12 \dow_\q v(t,\q)^2 + \frac{\pi^2}2 \dow_\q
\r(t,\q)^2 &= -W'(\q)
\end{split}
\een
where 
\be
v(t,\q) = \dow_\q \pi(t,\q).
\ee
These are coupled, non-linear partial differential equations and hence it is difficult to find a solution in general. One can decouple these two equations by introducing two new variable $p_\pm(t,\theta)$
\be\label{eq:BFdic}
\r(t, \q) ={p_+(t,\q)-p_-(t,\q)\over 2\pi}, \quad  \tand \quad 
v(t,\q) ={p_+(t,\q)+p_-(t,\q)\over 2}.  
\ee
The equations for $p_\pm(t,\theta)$ become
\ben\label{eq:fermisurfeom}
\dow_t p_{\pm}(t,\q) + p_{\pm}(t,\q)\dow_\q p_{\pm}(t,\q) +W'(\q) =0.
\een
The set of decoupled equations (\ref{eq:fermisurfeom}) governs the evolution of a free-Fermi droplet in $(p,\theta)$ plane whose boundaries are given by $p_\pm(t,\theta)$ \cite{polchinski}. To understand this in detail, consider a system of $N$ free Fermions (non-interacting) moving on $S^1$ under a common potential $W(\theta)$. The single particle Hamiltonian is given by
\ben\label{eq:singlep}
\mathfrak{h}(p,\theta) = \frac{p^2}2 + W(\q).
\een
The Hamilton's equations obtained from the single particle Hamiltonian (\ref{eq:singlep}) are given by
\ben
\frac{dp}{dt} = -W'(\theta), \qquad \frac{d\theta}{dt} = p.
\een
Using these equations one can check that the boundaries of a droplet $p\ra p_\pm(t,\theta)$ follow equation (\ref{eq:fermisurfeom}). Therefore equations (\ref{eq:fermisurfeom}) determines classical evolution of Fermi surface with time. The phase space Hamiltonian for such free Fermi system is given by,
\be\label{eq:HamF}
H_p = \frac1{2\pi} \int d\q \ dp \lb \frac{p^2}2 + W(\q) \rb \varpi(p,\q)
\ee
where $\varpi(p,\q)$ is the phase space density
\ben
\varpi(p,\q) = \Theta\lb(p_+(t,\theta)-p)(p-p_-(t,\theta))\rb.
\een

There is a one to one correspondence between phase space variables and collective field theory variables. Eigenvalue density and the corresponding momentum can be obtained from phase space distribution by integrating over $p$
\ben \label{eq:BFdic2}
\r(t,\q) =\frac1{2\pi} \int dp \ \varpi(p, \q), \quad
v(t,\q) = \frac1{2\pi\r} \int dp \ p \ \varpi(p,\q). 
\een
Thus the relations (\ref{eq:BFdic}) serve as a dictionary between bosonic and fermionic (phase space) variables. Integrating over $p$ in (\ref{eq:HamF}) we have,
\ben \label{eq:Hp}
H_p = \frac{1}{2\pi}\int d\theta \lb \frac{p_+(t,\theta)^3}{6}+W(\theta) p_+(t,\theta)\rb - \frac{1}{2\pi}\int d\theta \lb \frac{p_-(t,\theta)^3}{6}+W(\theta) p_-(t,\theta)\rb.\nonumber\\
\een
Using the dictionary (\ref{eq:BFdic}) the Hamiltonian (\ref{eq:HamF}) reduces to Hamiltonian (\ref{eq:HamB}). Thus we see that the matrix model (\ref{eq:mmqmech}) has two equivalent descriptions.

Solving the field theory equations of motion (\ref{eq:colleom}) is equivalent to solving for upper and lower Fermi surfaces in phase space picture. In either case, one needs to provide an initial data on a constant time slice in $(t,\q)$ plane. After that the problem reduces to a {\it Cauchy problem}. Existence of a unique solution depends on the geometry of initial data curve\footnote{See \cite{pallab}, for an example.}.

For a unitary matrix model, phase space distribution has $\q\ra -\q$
symmetry. Using this fact it is possible to show that the phase space area
covered by Fermi surfaces is a constant of motion. Area covered by Fermi surfaces at a time $t$ is given by
\be
\cA(t) = \int_{-\q_0}^{\q_0} \lb p_+(t,\q) -p_-(t,\q)\rb d\q.
\ee
Using equation (\ref{eq:fermisurfeom}) and the fact that $p_\pm(t,\q)$ and $W(\q)$ are symmetric functions of $\q$ it is easy to verify that 
\be
\frac d{dt} \cA(t)=0.
\ee
Thus, phase space area is preserved during classical time evolution. One can normalise the area to be unity : $\cA(t)=1$. It is only the shape which changes during evolution.


\section{Quantisation of droplets}\label{sec:quantisation}

In order to regularise the total number of states in phase space we divide the phase space into unit cells with volume $\hbar$ such that
\ben
\frac{1}{2\pi \hbar}\int dp d\theta \varpi(p,\theta) = N, \quad \with \quad  \hbar N =1 .
\een
The classical limit corresponds to $N\ra \infty, \ \hbar\ra0 \ \with \  \hbar N =1$.
We also modify our phase space Hamiltonian (\ref{eq:Hp}) accordingly and is given by
\ben \label{eq:HpQ}
\begin{split}
H_p = & \frac{1}{2\pi\hbar}\int d\theta \lb \frac{p_+(t,\theta)^3}{6}+W(\theta) p_+(t,\theta)\rb \\
& - \frac{1}{2\pi \hbar}\int d\theta \lb \frac{p_-(t,\theta)^3}{6}+W(\theta) p_-(t,\theta)\rb.
\end{split}
\een

Our goal now is to find a simplectic form on the phase space  \cite{Grant:2005qc,Maoz:2005nk}, so that the Hamilton's equation
\be\label{eq:Hameq2}
\dot {p}_{\pm}(t,\theta) = \{p_\pm(t,\theta),H_p\}
\ee
coincides with (\ref{eq:fermisurfeom}), where $H_p$ is the total Hamiltonian (\ref{eq:HpQ}). To achieve this goal we introduce equal time Poisson brackets between $p_{\pm}(t,\theta)$ and $p_{\pm}(t,\theta')$
\ben\label{eq:poisson}
\{p_\pm(t,\theta),p_\pm(t,\theta')\} = \pm {2\pi \hbar} \delta'(\theta-\theta')\quad \text{and} \quad \{p_+(t,\theta), p_-(t,\theta')\} =0.
\een
It is easy to check that using these Poisson brackets the equation (\ref{eq:Hameq2}) boils down to (\ref{eq:fermisurfeom}).

\subsection{Quantisation and Kac-Moody algebra}\label{subsec:quantisation}

Suppose $p_\pm^{(0)}(t,\theta)$ are solutions of equation (\ref{eq:fermisurfeom}). Consider fluctuations $\tilde{p}_\pm(t,\theta)$ about such classical solutions
\ben
p_\pm(t,\theta) = p_\pm^{(0)}(t,\theta) + \hbar \ \tilde p_\pm(t,\theta).
\een
The fluctuations $\tilde{p}_\pm(t,\theta)$ satisfy
\ben\label{eq:fermisurfeomQ}
\dow_t \tilde p_{\pm}(t,\q) + \dow_\theta (p_\pm^{(0)}(t,\theta) \tilde p_{\pm}(t,\q)) + \hbar \ \tilde p_{\pm}(t,\q)\dow_\q \tilde p_{\pm}(t,\q) =0.
\een
This equation follows from the Hamiltonian
\ben\label{eq:hamfluctuation}
\begin{split}
    \tilde H_p & = \frac{1}{2\pi\hbar}\int d\theta \lb \frac{\hbar^3}{6} \tilde p_+(t,\theta)^3+ \frac{\hbar^2}{2}  p_+^{(0)}(t,\theta) \tilde p_+(t,\theta)^2\rb \\
    & \quad - \frac{1}{2\pi\hbar}\int d\theta \lb \frac{\hbar^3}{6} \tilde p_-(t,\theta)^3+ \frac{\hbar^2}{2}  p_-^{(0)}(t,\theta) \tilde p_-(t,\theta)^2\rb
\end{split}
\een
with the following Poisson bracket relations
\ben\label{eq:poissonq}
\{\tilde p_\pm(t,\theta), \tilde p_\pm(t,\theta')\} = \pm \frac{2\pi}{\hbar} \delta'(\theta-\theta')\quad \text{and} \quad \{\tilde p_+(t,\theta), \tilde p_-(t,\theta')\} =0.
\een
We also assume that the fluctuations preserve the total area of the droplet. This implies that 
\ben\label{eq:tildepcons}
\int_{-\pi}^{\pi} d\theta (\tilde p_+(t,\theta) - \tilde p_-(t,\theta))=0.
\een

The unitary matrix model (\ref{eq:mmqmech}) admits a minimum free energy classical configuration given by circular droplet $p_\pm^{(0)}(t,\theta) = \pm \frac12$ when $W(\theta)$ is constant. We study quantum fluctuations about this solution. However our analysis can be followed to study quantum fluctuations about other classical configurations as well. For $p^{(0)}_\pm =\pm \frac12$ the Hamiltonian (\ref{eq:hamfluctuation}) is given by
\ben\label{eq:Htilde}
\tilde H_p = \frac{\hbar}{8\pi}\int_{-\pi}^{\pi} \lb \tilde p_+^2(t,\theta) + \tilde p_-^2(t,\theta) \rb d\theta + \frac{\hbar^2}{12 \pi} \int_{-\pi}^{\pi} \lb \tilde p_+^3(t,\theta) - \tilde p_-^3(t,\theta) \rb d\theta.
\een
$\tilde p_\pm$ satisfies
\ben\label{eq:eomgs}
\dow_t \tilde p_{\pm}(t,\q) + \frac12 \dow_\theta \tilde p_{\pm}(t,\q) + \hbar \ \tilde p_{\pm}(t,\q)\dow_\q \tilde p_{\pm}(t,\q) =0.
\een
To quantise the above classical system we promote the Poisson brackets (\ref{eq:poissonq}) to commutation relations
\be\label{eq:commutation}
\left[ \tilde p_{\pm}(t,\theta), \tilde p_{\pm}(t,\theta')\right] = \pm 2\pi i \delta'(\theta-\theta')\quad \text{and}\quad [\tilde p_{+}(t,\theta),\tilde p_-(t,\theta')]=0.
\ee
We decompose $\tilde p_\pm(t,\theta)$ into Fourier modes
\ben\label{eq:hpmode}
\tilde p_{+}(t,\theta) = \sum_{n=-\infty}^\infty a_{-n}(t) e^{i n \theta}
\een
and
\ben\label{eq:hmmode}
\tilde p_{-}(t,\theta) =  -\sum_{n=-\infty}^\infty b_{n}(t) e^{i n \theta}.
\een
The constraint (\ref{eq:tildepcons}) implies that the zero-modes $a_0$ and $b_0$ are equal up to a sign
\ben\label{eq:abidentification}
a_0=-b_0=\pi_0.
\een
It follows from the quantisation conditions (\ref{eq:commutation}) that the Fourier modes $a_n$ and $b_n$ satisfy $u(1)$ \emph{Kac-Moody} algebra
\ben \label{eq:U1KM}
[a_m(t),a_n(t)]=m\delta_{m+n},\quad [b_m(t),b_n(t)]=m\delta_{m+n}, \quad \tand \quad [a_m(t),b_n(t)]=0.\ \ 
\een
The Hamiltonian (\ref{eq:Htilde}) in terms of these modes are given by
\ben \label{eq:Hamab}
\begin{split}
\tilde H_p = & \frac{\hbar}{4}\sum_n\lb a_n(t)a_{-n}(t)+b_n(t)b_{-n}(t)\rb\\
 & + \frac{\hbar^2}{6}\sum_{m,n}\lb a_m(t)a_n(t)a_{-(m+n)}(t) + b_m(t)b_n(t)b_{-(m+n)}(t) \rb
\end{split}
\een
The phase space Hamiltonian is not a free Hamiltonian, it contains a cubic interaction. Our next goal is to construct the Hilbert space for the system of quantised droplet and set up a map between different states in the Hilbert space and shapes of droplets (excitations).

Before constructing the Hilbert space we note that in $\hbar\ra 0$ limit the quantum excitations $\tilde p_\pm(t,\theta)$ can be related to primary fields associated with a theory of free bosons moving on a cylinder. We use the equations of motion (\ref{eq:eomgs}) in $\hbar\ra 0$ limit and find that the time evolution for $a_n(t)$ and $b_n(t)$ are given by
\ben
a_k(t) = a_k(0) e^{ikt/2}, \quad b_k(t) = b_k(0) e^{ikt/2}.
\een
Therefore $\tilde p_\pm$ can be written as (we define $a_n \equiv a_n(0)$ and $b_n \equiv b_n(0)$)
\ben
\tilde p_+(t,\theta) = \sum_n a_n e^{-n (\tau+i\theta)} \quad \tand \quad \tilde p_-(t,\theta) = \sum_n b_n e^{-n (\tau-i\theta)}\quad \where \ \tau = \frac{it}{2}
\een
Defining $z=e^{\tau +i\theta}$ and $\bar z = e^{\tau-i \theta}$ we see that $\tilde p_+$ is a holomorphic function of $z$ and $\tilde p_-$ is an anti-holomorphic function of $\bar z$
\ben\label{eq:pexpansion}
\tilde p_+(z) = \pi_0 + \sum_{k\neq 0} a_k z^{-k} \quad \text{
and} \quad \tilde p_-(\bar z) = \pi_0 + \sum_{k\neq 0} b_k \bar z^{-k}
\een
and the modes satisfy $u(1)$ Kac-Moody algebra. Thus, in the $N\ra \infty$ limit $p_+$ and $p_-$ are related to holomorphic and anti-holomorphic conserved currents associated with a free scalar CFT on a cylinder of radius \emph{one}\footnote{From the above construction it is easy to see that the fluctuations $\tilde p_\pm$ can be related to conserved currents associated with a free scalar on a cylinder of radius \emph{one}. We define a scalar field on a cylinder of radius \emph{one}
\ben
\varphi(z,\bar z) = \varphi_0 - i \pi_0 \ln(z\bar z) + i \sum_{n\neq 0} \frac{1}{n}\lb a_n z^{-n}+b_n \bar z^{-n}\rb
\een
such that
\ben
i \partial \varphi = \frac{\pi_0}{z} + \sum_{n\neq 0} a_n z^{-n-1} = \frac{\tilde p_+(z)}{z},\quad 
\tand \quad
i \bar \partial \varphi = \frac{\pi_0}{\bar z} + \sum_{n\neq 0} b_n \bar z^{-n-1} = \frac{\tilde p_-(\bar z)}{\bar z}.
\een
}. 

The Hamiltonian $\tilde H_p$ also separates into two parts : $\tilde H_p =  H_+ +H_-$. Since the Hamiltonian (\ref{eq:Hamab}) does not depend on time explicitly, $ H_\pm$ are given by (up to an overall constant) 
\ben\label{eq:Hamat}
\begin{split}
    H_+ = & \overbrace{\frac{\hbar}{4}a_0^2 -\frac{\hbar^2}{24}a_0 + \frac{\hbar^2}{6}a_0^3}^{\displaystyle{H^+_0}}\\ & +\underbrace{ \frac{\hbar}2 \lb 1+2\hbar a_0\rb \sum_{k> 0}a^\dagger_ka_k }_{\displaystyle{H^+_{\text{free}}}}+ \underbrace{\frac{\hbar^2}{2} \sum_{m,n>0} \lb a^\dagger_{m+n}a_ma_n + h.c.\rb}_{\displaystyle{H^+_{\text{int}}}}
\end{split}
\een
and similarly
\ben\label{eq:Hbmat}
\begin{split}
    H_- = \frac{\hbar}{4}b_0^2 -\frac{\hbar^2}{24}b_0 + \frac{\hbar^2}{6}b_0^3 +\frac{\hbar}2 \lb 1+2\hbar b_0\rb \sum_{k> 0}b^\dagger_k b_k + \frac{\hbar^2}{2} \sum_{m,n>0} \lb b^\dagger_{m+n}b_mb_n + h.c.\rb.\ \ \ \ \ 
\end{split}
\een
Both the Hamiltonians have two parts, a free part and an interacting part. This Hamiltonian is similar to the Hamiltonian obtained by \cite{Minahan:1993np,polchinski,Douglas:1993wy} for splitting-joining of strings. Here the interaction piece is responsible for the interaction between the boundary excitations.

\subsection{The Hilbert space }\label{sec:phasestate}

We now construct the Hilbert space for `$+$' sector. There exists an isomorphic Hilbert space for `$-$' sector. We denote these Hilbert spaces by $\cH_\pm$ and the total Hilbert space is therefore given by $\cH=\cH_+\otimes\cH_-$.

Since $\pi_0$ commutes with all the $a_n, b_n$ and hence with $\tilde H_p$, application of $a_n$'s and $b_n$'s cannot change the eigenvalue of $\pi_0$. Therefore the Hilbert space is constructed upon a one parameter family of vacua $\ket{s,s}\equiv \ket s$ where,
\ben
\label{eq:KMprimary}
\begin{split}
   & a_n\ket{s} = 0, \quad b_n\ket{s} = 0 \quad \for\  n>0\\
\tand \quad & a_0\ket{s} = - b_0\ket{s} =\pi_0\ket{s} = s \ket{s}.
\end{split}
\een
The Hilbert space constructed upon $\ket s$ vacuum is denoted by $\cH$ : $s$ charged module generated over the primary $\ket{s}$.  We take $s$ to be integer in order to get the phase space momentum (\ref{eq:BFdic2}) quantised in the units of $\hbar$.

A generic excitation above the ground state is given by
\ben\label{eq:basisstatek}
\ket{\vec k,\vec l} = \prod_{n,m=1}^{\infty} \lb a^\dagger_n\rb^{k_n}\lb b^\dagger_m\rb^{l_m}\ket{s}.
\een
The $\vec k$ and $\vec l$ sectors correspond to excitations in upper and lower Fermi surfaces. Since $a_n$ and $b_n$ commute, generic excitation $\ket{\vec k, \vec l} \in \cH$ can be written as $\ket{\vec k} \otimes \ket{\vec l}$. The time evolution of the vectors in $\cH_\pm$ are governed by $H_\pm$ respectively and are independent, except for the fact that $ p_+- p_- \geq 0$. We first consider the states and their evolution in $\cH_+$ only. $\cH_-$ can be studied similarly\footnote{One can study an entangled excitations of free Fermi droplets. We do not discuss such excitations in this paper.}. Later in section \ref{sec:combinetwosector} we combine the evolution of classical states in these two sectors and show how they are related to correlation functions of $2d$ Yang-Mills theories on Riemann surfaces.

The excited states in $\cH_+$ are orthogonal with the normalization
\ben\label{eq:knormalization}
\braket{\vec{k'}}{\vec k} =z_{\vec k}\delta_{\vec k \vec{k'}}\ \ \text{where}\ z_{\vec{k}}=\prod_j k_j! j^{k_j}
\een
and has $\pi_0$ eigenvalue $s$. They also satisfy the completeness relation
\ben
\sum_{\vec k}\frac{1}{z_{\vec k}} \ket{\vec k}\bra{\vec k} = {\mathbb{I}}_{\cH_+}
\een
and hence form a basis in $\cH_+$. These states are particle like excitation above the ground state $\ket s$. The  excited states $\vec k$ in either sectors are eigenstates of the free Hamiltonian
\ben
H^\pm_{\text{free}}\ket{\vec k} = \frac{\hbar}{2}(1 \pm 2s\hbar)\lb\sum_{n=1}^\infty n k_n\rb \ket{\vec k},
\een
but not an eigenstate of the full Hamiltonian. The interaction Hamiltonian changes $\ket{\vec k}$ state to $\ket{\vec k'}$ keeping the level fixed, \emph{i.e.} $\sum_n n k_n =\sum_n nk_n'$.
The expectation value of $p_+(t,\theta)$ operator in $\ket{\vec k}$ state is $(1/2+\hbar s) z_{\vec{k}}$ with a non-zero quantum dispersion $\Delta p_+$ which goes as $\hbar$. Therefore $\ket{\vec k}$ states are quantum excitations over the ground state : ripples on Fermi surface. In \cite{Minahan:1993np} such excitations in $2d$ Yang-Mills were identified with the left and right winding of strings around the circle - excitation of $k_n$ closed strings winding $n$ times around the circle.

One can also define coherent state in $\cH_+$
\ben\label{eq:coherentstate}
\ket{\t_+} = \exp\lb {\sum_{n=1}^\infty \frac{\t_{n}^+ a_n^\dagger}{n\hbar}}\rb \ket{s}.
\een
The state $\ket{\t_+}$ is not normalised, one can show that
\ben
\braket{\tau_+^a}{\tau^b_+} = \exp\lb\sum_n \frac{\t_+^a\t_+^b}{n\hbar^2}\rb.
\een
The coherent state $\ket{\t_+} $ is an eigenstate of $a_n$ ($\forall\, n>0$) operator with eigenvalue $\t_{n}^+/\hbar$. A coherent state $\ket{\t_+}$ can be expanded in $\ket{\vec k}$ basis in the following way
\ben
\ket {\t_+} = \sum_{\vec k} \frac{\t^+_{\vec k}}{z_{\vec k}}\ket{\vec k}, \quad \where \quad \t^+_{\vec k} = \prod_m \lb\frac{\t^+_{m}}{\hbar}\rb^{k_m}.
\een
The expectation value of $p_+$ operator in a coherent state $\ket {\t_+}$ is given by
\ben
\omega_{\t_+}(z)= \frac{ \bra{\t_+}\frac{p_+(z)}{2\pi} \ket{\t_+}}{\braket{\t_+}{\t_+}} =\frac{1}{4\pi} + \frac{s\hbar}{2\pi}+\frac{1}{2\pi}\sum_{n>0}\t^+_{n}\lb z^n +\frac{1}{z^n}\rb.
\een
Value of $\omega_{\t_+}(z)$ on the unit circle ($|z|=1$) in the complex $z$ plane is given by, 
\ben\label{eq:coherentprofile}
\begin{split}
\omega_{\t_+}(\theta) \equiv \omega_{\t_+}(z=e^{i\theta}) & = \frac{1}{4\pi}  + \frac{s\hbar}{2\pi} + \tilde{\omega}_{\t_+}(\theta)\\
\text{where} \quad \tilde{\omega}_{\t_+}(\theta) &=\frac{1}{\pi} 
\sum_{n>0}\t^+_{n} \cos n\theta.
\end{split}
\een
The quantum dispersion of $p_+$ in a coherent state is zero. Therefore such states are called \emph{classical}. 

We now define the following mapping between a state $\ket{\psi} \in \cH_+$ and shape of the upper Fermi surface
\ben\label{eq:maping}
\ket{\psi} \ra \{\bra{\psi}p_+(\theta)\ket{\psi}\}.
\een
This mapping maps the following three types of states in $\cH_+$ to three different types of shapes of the upper Fermi surface $p_+$.
\begin{itemize}
    \item Expectation value of $p_+$ in ground state is $\bra{s}p_+ \ket{s} = \frac12 + s \hbar$ with zero dispersion. Therefore the ground state $\ket s$ corresponds to an overall shift of $p_+$ by an amount $s/N$ over the classical value. Similar mapping exists in the $-$ sector as well and the ground state corresponds to the same shift in $p_-$. Since the eigenvalue density is proportional to the difference between $p_+$ and $p_-$, such constant shift in upper and lower Fermi surfaces does not change the eigenvalue distribution. \item Expectation value of $p_+$ in a generic \emph{normalised} excited  state is same as the expectation value in the ground state with non-zero dispersion\footnote{This is similar to the expectation of position or momentum operator of a simple harmonic oscillator in the ground state or higher excited states.}. Therefore excited states correspond to ${\cal O}(\hbar)$ ripples on $p_+$.
    \item The relation (\ref{eq:coherentprofile}) defines a mapping between a coherent states $\ket{\t_+}$ in $\cH_+$ and a classical deformation of droplet in `$+$' sector over the ground state. Specifying a coherent state is equivalent to specifying a ${\cal O}(1)$ deformation $\o_{\t_+}(\theta)$ of $p_+$. The mapping is one-to-one as long as a classical distribution $\bra{\psi}p_+(\theta)\ket{\psi}$ is a single valued function of $\theta$. 
\end{itemize} 
An important thing to note here is that classical deformations do not disturb the quadratic profile of the droplets \emph{i.e} for a given $\theta$, there exists unique values of $p_\pm(\theta)$.
\begin{figure}[h]
	\begin{center}
		\includegraphics[width=5cm,height=6cm]{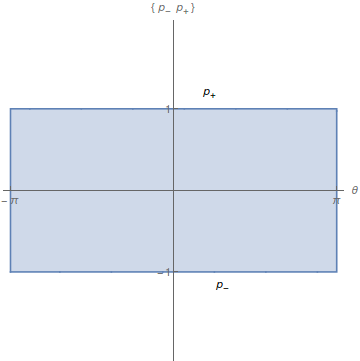}
		\includegraphics[width=5cm,height=6cm]{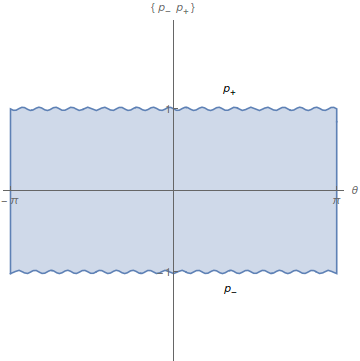}
		\includegraphics[width=5cm,height=6cm]{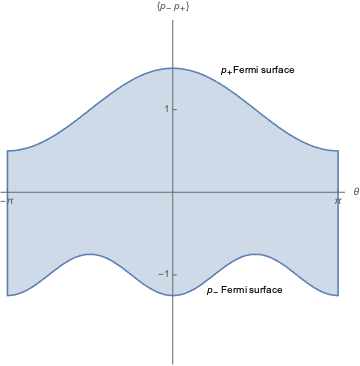}
		\caption{\footnotesize{Different Fermi surfaces.}}
		\label{fig:fermisurfaces}
	\end{center}
\end{figure}
There are other types of excitation, which destroy the quadratic profile of a droplet - formation of folds. We shall discuss about fold states in a future work.

\subsection{Eigenstates}
\label{sec:eigenstate}

The basis states $\ket{\vec k}$ in $\cH_+$ are not eigenstates of the full Hamiltonian $H_+$. We introduce a new basis in the Hilbert space $\cH_+$ - \emph{Young diagram basis}. We associate a state $\ket {R_+}$ in $\cH_+$ for a given Young diagram $R_+$ in the following way
\ben\label{eq:YDbasis}
\ket {R_+} = \sum_{\vec k}\frac{\chi_{R_+}(\vec k)}{z_{\vec k}} \ket{\vec k}
\een
where $\chi_{R_+}(\vec{k})$ is the character of the conjugacy class $C(\vec k)$ of the permutation group $S_K$. $\ket{R_+}$ corresponds to a Young diagram $R_+$ whose rows have lengths $l_1\geq l_2\geq l_3\geq \cdots\geq l_r\geq 0$, with $r$ being the number of rows. Total number of boxes in $R_+$ is equal to the level of $\ket{\vec k}$ states : $\sum_i n_i =\sum_n nk_n$. Following the normalization (\ref{eq:knormalization}) of $\ket{\vec k}$ we have %
\be
\braket{R_+}{R_+'}=\delta_{R_+ R_+'}.
\ee
Inverting the relation (\ref{eq:YDbasis}) we have
\ben
\ket{\vec k} = \sum_{R_+} \chi_{R_+}(\vec k) \ket {R_+}.
\een
Thus $\ket{\vec k}$ and $\ket{R_+}$ are two equivalent basis of the Hilbert space $\cH_+$. The young diagram basis also has interpretations in terms of fermionic excitations following bosonisation \cite{Douglas:1993wy,Marino:2005sj}.

It turns out that (see appendix \ref{app:eigenphase}) the Hamiltonian (\ref{eq:Hamat}) is diagonal in the Young diagram basis
\begin{equation}\label{eq:Reigenstate}
   H_+\ket {R_+} = \Big[{\hbar^2\over 6}s^3+{\hbar\over 4}s^2-{\hbar^2\over 24}s+{\hbar\over 2}\left(1+2s\hbar\right)l(R_+)+{\hbar^2\over 2}\kappa_{R_+}\Big] \ket {R_+} \equiv E(R_+,s)\ket{R_+}
\end{equation}
where
\ben
l(R_+)=\sum_{i=1}^r l_i\quad \tand \quad   \kappa_{R_+}=l(R_+)+\sum_{i=1}^r (l_i^2-2il_i).
\een
Note that eigenvalues of the free and the interaction part of the Hamiltonian are of the same order $\hbar l(R_+) \sim \hbar^2 \kappa_{R_+}$ in the large $N$ limit. Therefore, the cubic part (interaction) can not be treated as perturbation in the large $N$ limit.

The quadratic Casimir $C_2(R_+)$ of $u(N)$ representation is given by
\ben
    C_2(R_+) &=N\,l(R_+)+\kappa_{R_+}.
\een
From equation (\ref{eq:Reigenstate}) we see that in the limit $N\ra \infty$ the eigenvalue $E(R_+,s)$ of $\ket{R_+}$ is equal to quadratic Casimir $C_2(R_+)$ for $R_+$ representation
\ben\label{eq:Reigenvalue}
E(R_+,s) = \frac{\hbar^2}{2} C_2(R_+) +\cO(\hbar).
\een
The Hilbert space, discussed above, is similar to the Hilbert space of a chiral sector of $U(N)$ $2d$ Yang-Mills on Riemann surfaces\footnote{The complete Hamiltonian can be obtained by considering excitations in both $p_+$ and $p_-$ sectors.}. The states (\ref{eq:basisstatek}) in $\cH_+$, with all $l_m=0$ are the states of interacting strings that wind around the circle in one direction. The Hamiltonian $H_+$ is thus the Hamiltonian for the closed strings dual to the chiral sector of two dimensional Yang-Mills \cite{Minahan:1993np,Minahan:1993tp,Grossmatytsin, Gross:1993hu,Douglas:1993wy,Donnelly:2016jet}. The evolution of the upper Fermi surface $p_+$, therefore, maps to dynamics of string degrees of freedom dual to the chiral sector of $2d$ Yang-Mills. The evolution of lower Fermi surface $p_-$, in a similar way, provides the other chiral sector. 

\section{Evolution of classical droplets}
\label{sec:correlation}

In quantum mechanics all the information of a system is stored in the propagator : the transition amplitude from an initial state $\ket{i}$ at $t_i =0$ to a final state $\ket{f}$ at time $t_f = T$, given by
\ben\label{eq:ampli-f}
K_{i\ra f}(T) = \bra{f}e^{\frac{i}{\hbar}H T}\ket{i}.
\een
The Euclidean version of this amplitude is obtained by replacing $T\ra -i T $
\ben\label{eq:Eampli-f}
\cK_{i\ra f}(T) = \bra{f}e^{-\frac1\hbar H T}\ket{i}.
\een
We define $\mathbb{ S}_+$ : a set of all classical droplet configurations over the ground state in $\cH_+$. There exists a one-to-one mapping between $\mathbb{ S}_+$ and the coherent states of $\cH_+$. Our goal is to compute the propagators\footnote{
If we take the initial state to be a representation $\ket {R}$, it will remain in the same state as $\ket {R}$ is an eigenstate of $H_+$. A transition between initial state $\ket{\vec k}$ and and final state $\ket{\vec l}$ is given by
\ben
\cK(\vec k\ra \vec l; T) = \sum_R \chi_R(\vec k)\chi_R(\vec l) e^{-\frac{\hbar}{2}E(R)T}.
\een
This amplitude is zero if the levels of $\ket{\vec k}$ and $\ket{\vec l}$ are different.} in $\mathbb{ S}_+$.

In order to calculate the transition amplitude between  coherent states $\ket{\t_+^a}$ and $\ket{\t_+^b}$ in $\mathbb{ S}_+$ we first expand a coherent state in the Young diagram basis
\ben\label{eq:tauinR}
\ket {\t_+} = \sum_{R_+} \lb \sum_{\vec k} \frac{\chi_{R_+}(\vec k)}{z_{\vec k}} \t^+_{\vec k}\rb \ket{R_+} .
\een
The term inside the parenthesis can be simplified using equation (\ref{eq:coherentprofile}). $\t^+_k$ is given by
\be
\t^+_k = \int_{-\pi}^{\pi}d\theta \ \tilde\omega_{\t_+}(\theta) e^{i k \theta}.
\ee
Since $\tilde\o_{\t_+}(\q)$ defines a distribution of $\theta_i$ in $N\ra \infty$ limit, we can write 
\ben
\tilde \o_{\t_+}(\theta) = \lim_{N\ra \infty} \frac1N \sum_{i=1}^N \delta(\theta-\theta_i) -\frac{1}{2\pi}.
\een
Hence, we have
\be\label{eq:mapcohtoholo}
\frac{\t^+_k}{\hbar} = \sum_{i}e^{i k \theta_i}. 
\ee
This equation provides a map between infinite dimensional $\t^+ =\{\tau^+_1,\tau^+_2,\cdots\}$ space and the set of angular variables $\{\theta_1, \cdots, \theta_N\}$ in $N\ra \infty$ limit. This mapping is one to one - a coherent state $\ket{\tau_+} \equiv \{\tau^+_n\}$ corresponds to a unique point $\{\theta_i\}$ in $\theta$ space. Using (\ref{eq:mapcohtoholo}) we can write
\ben\label{eq:tauschur}
\sum_{\vec k} \frac{\chi_{R_+}(\vec k)}{z_{\vec k}} \t^+_{\vec k} =  \Tr_{R_+} U(\infty) \equiv s_{R_+}(\t_+),
\een
where $s_{R_+}(\t_+)$ is Schur polynomial\footnote{In equation (\ref{eq:tauschur}), Schur polynomial $s_{R_+}$ is a function of $\theta_i$s. Since equation (\ref{eq:mapcohtoholo}) maps $\t^+_i$ space to $\theta_i$ space we write $s_{R_+}$ as a function of $\t_+$.}. Thus we have
\ben\label{eq:tausR}
\ket{\tau_+} = \sum_{R_+} s_{R_+}(\tau_+) \ket {R_+}.
\een
From the closed string point of view $s_{R_+}(\t_+)$ is the wave function of chiral winding string states in $\ket{R_+}$ basis \cite{Donnelly:2016jet}. A coherent state in $\cH_+$ captures information of winding states in the chiral sector of closed string theory. Thus a classical shape of upper Fermi surface has a correspondence with winding states in the chiral sector of closed strings dual to $2d$ Yang-Mills.

We now consider the transition amplitude between two coherent states $\ket{\t^a_+}$ and $\ket{\t^b_+}$ in time interval $T$
\ben\label{eq:Eampli-t}
\cK_+(a\ra b,T) = \bra{\t^b_+}e^{-\frac1\hbar H_+ T}\ket{\t^a_+}.
\een
Using relation (\ref{eq:tausR}) we can express this transition amplitude as a sum over representations
\ben\label{eq:cylinder}
\cK_+(a\ra b,T) = \sum_{R_+} s_{R_+}(\t^a_+)s_R(\t^b_+) e^{- \frac{\hbar}2 C_2(R_+)T}.
\een
The transition amplitude between two coherent states in time $T$ is same as the chiral partition function of $2d$ Yang-Mills theory on a cylinder with holonomies specified at the two circular ends \cite{Gross:1993yt,Gross:1993hu,Gross:1992tu,Grossmatytsin}. The droplet profiles of the initial and the final coherent states are mapped to these two holonomies. We denote such an amplitude by $\cC_+(a, b,T)$.

There is a special point in $\tau$ space, $\t^* : \t^+_i=1$, $\forall i$. Droplet profile for such a coherent state is given by
\ben
\omega_{\tau^*}(\theta) =  -\frac{1}{4\pi}+\frac{\hbar s}{2\pi}+\delta(\theta).
\een
Schur polynomial for such a distribution is equal to the dimension of the representation $R_+$, denoted by $\dim R_+$. If the initial (or final) configuration corresponds to this particular configuration then $\cC_+(\t^* \ra b,T)$ is given by
\ben
\cC_+(\t^*,b,T) = \sum_{R_+} s_{R_+}(\t^b) \text{dim}R_+ \ e^{- \frac{\hbar }2 C_2(R_+)T}.
\een
The delta function distribution of $\omega_{\tau}(\theta)$ at one end of the cylinder is equivalent to shrinking the radius of that end to \emph{zero}. As a result the corresponding amplitude becomes a disk amplitude $\cD_+(b,T)\equiv \cC_+(\t^*,b,T)$. For both $\t_+^a=\t_+^b = \t^*$, we get
\ben\label{eq:sphere}
\cC_+(\t^*, \t^*,T) = \sum_{R_+} \lb \text{dim}R_+ \rb^2 e^{- \frac{\hbar}2 C_2(R_+)T}.
\een
This is the chiral partition function of $2d$ Yang-Mills on a sphere.

We next define a surgery in $\tau_+$ space to generate higher point function. Consider a cylinder $\cC_+(a,b,A_1)$ and a disk $\cD_+(c,A_2)$. To join the outer edge of the disk with one side (say the $b$ side) of the cylinder we first set $\tau_+^c=\tau_+^{b*}$. Then sum over all possible coherent states $\ket{\tau^b_+}$. Since $\t_+$ space is isomorphic to $\theta$ space, the integration in $\t_+$ space can be replaced by an integration in $\theta$ space with  the Vandermonde factor
$V(\theta_i)=\prod_{i<j}\sin^2{\frac{\theta_i-\theta_j}{2}}$ : 
\ben\label{eq:surgery}
\int [d\t_+] s_{R_+}(\t)s_{R'_+}(\t) \ra \int [d\theta] \prod_{i<j}\sin^2{\frac{\theta_i-\theta_j}{2}} s_{R_+}(\theta)s_{R'_+}(\theta) = \delta_{R_+R'_+}.
\een
Therefore we get,
\ben\label{eq:surgeryCD}
\int [d\t^b_+] \cC_+(a,b,T_1)\cD_+(b,T_2) = \cD_+(a,T_1+T_2). 
\een
Following the surgery one can also find a torus partition function of $2d$ Yang-Mills in the chiral sector
\ben
\cZ = \int d\t_+^a \cC_+(a,a,A) = \Tr e^{-T H_+}.
\een

\subsection{Higher point functions}
\label{sec:higherpointfunc}

In $2d$ Yang-Mills theory one can define higher point functions : partition function with more than two holonomies. Such partitions are mapped to a Riemann surface with more than two punctures. 

We define an operator $\cO_+(\t_+)$ associated with a coherent state $\ket{\t_+}$ in $\cH_+$
\ben\label{eq:cohop}
\cO_+(\t)= \sum_{R_+} \frac{\chi_{R_+}(\t_+)}{\mathrm{dim}R_+} \ket{R_+}\bra{R_+}.
\een
This operator has the properties 
\ben
\begin{split}
     & (a) \ \ \cO_+(\tau^*) =\mathbb{I}, \hspace{1.26cm}
    (b) \ \ [\cO_+(\tau_+^a),\cO_+(\tau_+^b)]=0,\\
    & (c) \ \ [\cO_+(\tau_+^a),H]=0, \quad (d) \ \ \cO_+(\tau_+)\ket{\tau^*}=\ket{\tau_+}.
\end{split}
 \een
Following (\ref{eq:coherentprofile}), the operator (\ref{eq:cohop}) defines a correspondence between classical shape $\tilde \omega(\t_+)$ and operators $\cO_+(\t)$
\ben
\tilde \omega(\t_+) = \bra{\t^*}\cO_+(\tau_+)\frac{\hat p_+(z)}{2\pi} \cO_+(\t_+)\ket{\t^*}\bigg|_{z=e^{i\theta}}.
\een

We introduce a density matrix for a mixed ensemble $\cE^+_g$ in $\cH_+$ for $g=0,1,2, \cdots$ 
\ben
\r^+_g=\sum_{R_+} \lb \mathrm{dim}R_+\rb^{2-2g} e^{-\frac{\hbar}{2}C_2(R_+)T}\ket{R_+}\bra{R_+}.
\een
The partition function in $\cE^+_g$ is given by
\ben
\cZ^+_g = \Tr \r^+_g = \sum_{R_+} \lb \mathrm{dim}R_+\rb^{2-2g} e^{-\frac{\hbar}{2}C_2(R_+)T}.
\een
Ensemble average of any operator $\cA$ is defined as 
$$\langle \cA \rangle_g=\Tr(\cA \r_g).$$
Therefore the ensemble average of a product of $n$ operators $\cO_+(\tau_+^{a_1})$, $\cdots$, $\cO_+(\tau_+^{a_n})$ is given by
\ben
\begin{split}
\langle \cO_+(\t_+^{a_1})\cdots \cO_+(\tau_+^{a_n}) \rangle_g &= \Tr\left ( \cO_+(\t_+^{a_1})\cdots \cO_+(\tau_+^{a_n})\r^+_g\right)\\
&=\sum_{R_+} \lb \mathrm{dim}R_+\rb^{2-2g -n} e^{-\frac{\hbar}{2}C_2(R_+)T} \prod_{i=1}^n s_{R_+}(\t_+^{a_i}).
\end{split}
\een
For $g=0$, $n=1$ and $n=2$ we get back our disc and cylinder amplitude respectively. $n=0$ gives the partition function of $2d$ Yang-Mills on a generic Riemann surface in the chiral sector. The operator $\cO_+(\tau_+)$, which corresponds to a classical distribution of droplet, creates a puncture on the Riemann surface. Three (and higher) point correlators in $\cH_+$ can also be thought of as evolution of initial droplet to a final one in the presence of some external disturbances in between. Such external disturbances can be mathematically expressed as local twists. For example, a $3$-point correlator can be interpreted as twisted surgery between two cylinders. See appendix \ref{app:higherpoint} for details.

\subsection{Connection to \emph{q}-deformed theories}\label{sec:q-def}

There exists another interesting class of coherent states given by \cite{Marino:2005sj}
\begin{equation}
\label{eq:qdefcoherent}
    \ket{\tau_{+}^q} =\exp\left[{\sum_{n=1}^{\infty}\frac{\g^{\frac n2}-\g^{-\frac n2}}{n(q^{\frac n2}-q^{-\frac n2})}a_n^{\dagger}}\right]\ket{s}
\end{equation}
where $q$ is a deformation parameter such that $0<q<1$ and $\g =q^{N}$. In the limit $q\ra 1$ we find that
\be
\lim_{q\ra 1}\ket{\t_{+}^q} =\exp\left[\sum_{n=1}^{\infty}\frac{1}{n\hbar}a_n^{\dagger}\right]\ket{s}=\ket {\t^*}\ .
\ee
The \emph{q}-deformed state $\ket{\t_{+}^q}$ has the property
\ben
\braket{R_+}{\t_{+}^q} = \text{dim}_q R_+ =\prod_{1\leq i<j\leq N}\frac{[h_i-h_j]}{[j-i]}
\een
where $\text{dim}_qR_+$ is \emph{q}-deformed dimension of $R_+$. Therefore the \emph{q}-deformed state can be written as
\ben
\label{eq:qdeformedstateinR}
\ket{\tau_+^q} =\sum_{R_+} \lb \text{dim}_q R_+\rb \ket{R_+}.
\een

Using the state droplet mapping one can find the shape of droplet for a \emph{q}-deformed coherent state. We take the deformation parameter $q = e^{i g_s}$ and then consider a double scaling limit $g_s\ra 0, \ N\ra \infty$ keeping $g_s N=\l$ fixed. In this limit $ \omega_{\t_{+}^q}(\theta)$ is given by\footnote{Here we use the following definition of \emph{q}-deformation
\ben
[x]=q^{x/2}-q^{-x/2}.
\een
}
\ben
\omega_{\t_{+ }^q}(\theta) =-\frac{1}{4\pi}+\frac{\hbar s}{2\pi} + \frac{1}{\lambda}\Theta\left(\frac{\lambda^2}{4}-\theta^2\right).
\een
The $q\ra 1$ limit corresponds to $\l\ra 0$. In this limit, the theta function approaches to $\delta(\theta)$, as expected.

The Lorentzian amplitude (\ref{eq:ampli-f}) from a $\ket{\t_{+}^q}$ state to a $\ket{\tau^b_+}$ state is given by\footnote{We consider Lorentzian amplitude since we have taken $q=e^{i g_s}$.}
\ben\label{eq:qddisc}
\cK(\t_{+}^q\ra \t^b_+,T) =\sum_{R_+} s_{R_+}(\t^b_+) \textrm{dim}_q R_+ \  q^{\frac12 C_2(R_+)}\quad \text{where $\hbar T=g_s$.}
\een
This amplitude is same as the disc amplitude of a $q$-deformed Yang-Mills theory \cite{Arsiwalla:2005jb}. This is also similar to the character expansion of a ``lesser known" generalized $U(N)$ Villain action (\ref{eq:unvilchar}) (up to a normalization factor) \cite{Onofri:1981qk, Romo_2012}. We have given a detailed discussion on Villain action in appendix \ref{app:villain}. In $\cH_+$, one can also consider an evolution of a coherent state to another coherent state : $\ket{\t_{+}^{ q_1}} \ra \ket{\t_{+}^{ q_2}}$.  In the droplet picture such an evolution corresponds to a box distribution evolving to another box distribution keeping the area preserved. The amplitude for such evolution is given by
\ben\label{eq:q1q2transition}
\sum_{R_+} \text{dim}_{q_{1}}R_+ \text{ dim}_{q_{2}}R_+ \hspace{3pt} q_1^{\frac{1}2 C_2(R_+)} q_2^{\frac{1}2 C_2(R_+)}.
\een
Such a propagator appears when we glue two Villain actions with different 't Hooft coupling $\lambda_1$ and $\lambda_2$ such that $\lambda_1+\lambda_2 = T$. The amplitude (\ref{eq:q1q2transition}) can be thought of as gluing two different \emph{q}-deformed disc over the edges. In (\ref{eq:q1q2transition}) if we take $q_1=1$ (or $q_2=1$) we get 
\ben\label{eq:q11q2transition}
\sum_{R_+} \text{dim}R_+ \text{ dim}_{q_{2}}R_+ \hspace{3pt} q_2^{\frac{1}2 C_2(R_+)}.
\een
This is the transition amplitude between  $\ket{\t_+^{q_2}}$ and $\ket{\tau^*}$. In the large $N$ limit these amplitudes give rise to a mixed Riemann-Hilbert problem \cite{riemann-hilbert-book}. Such Riemann-Hilbert problems appear in different contexts both in physics and mathematics. In physics, for example, they appear in open topological-A theory amplitudes on some specific Calabi-Yau caps \cite{Bryan:2004iq} and thus in the context of black hole microstate counting in type II string theory owing to the Ooguri-Strominger-Vafa (OSV) conjecture \cite{Ooguri_2004}. In mathematics, the Riemann-Hilbert problem associated with the \emph{q}-deformed Plancherel growth belong to the same class \cite{2007arXiv0706.3292S}.

We see that the \emph{q}-deformation in the Yang-Mills side is related to special deformation of droplet geometries without deforming the gauge group associated with the matrix model. Thus the geometry of droplet unifies different versions of $2d$ Yang-Mills theories. A droplet contains more information than it is expected.

\subsection{Joining two chiral sectors}
\label{sec:combinetwosector}

The Hilbert space $\cH_+$ captures only one chiral sector of the Yang-Mills theory on a circle \cite{Gross:1993hu}.
The propagator defined in \eqref{eq:cylinder}, gives only the chiral partition function of the $2d$ Yang-Mills theory on cylinder. In order to recover the full partition function, one has to consider the contribution of propagators coming from states in both the Hilbert spaces $\cH_+$ and $\cH_{-}$. Tools for factorizing Yang-Mills theories into chiral and anti-chiral sectors has been developed and discussed previously \cite{Gross:1993hu,Aganagic:2005dh,Donnelly:2016jet} using the notion of \emph{composite representations} (we elaborate on this further in appendix \ref{app:repgyan}). A composite representation $R_+\bar R_{-}$ can be built from the Young diagrams of $R_+$ and $R_{-}$ by drawing the diagram corresponding to $R_+$ in a standard way on top right and the Young diagram of $R_{-}$ as \emph{anti-boxes} in the bottom left corner turning it upside down. The remaining rows have zero boxes such that the total number of rows in the composite diagram is $N$. Note that such a procedure makes sense where each Young diagram $R_+$ and $R_{-}$ has less than $\frac{N}{2}$ rows. However, in $N\ra \infty$ limit one can extrapolate this procedure for any $R_+$ and $R_-$. The representations corresponding to the Young diagram of $R_+\bar{R}_{-}$ (as described above) forms a basis for $U(N)$. Another basis for $U(N)$ is given by $R_{+}\otimes R_{-}$. The relation between the two basis is given by
\begin{equation}\label{eq:composite-basischange}
    \ket{R_{+}}\otimes \ket{\bar R_{-}}=\sum_{S_+,S_{-}}\left( \sum_{S'}N^{R_+}_{S_+ S'}N^{R_{-}}_{S_{-}S'}\right)\ket{S_{+}\bar{S}_{-}}
\end{equation}
and the inverse relation
\ben\label{eq:composite-basischange-inverse}
\ket{S_{+}\bar{S}_{-}} = \sum_{R_+,R_-}\lb \sum_{R'}(-1)^{|R'|}N^{S_+}_{R_+R'}N^{S_-}_{R_-\bar R'}\rb \ket{R_+}\otimes\ket{\bar R_-}.
\een

A generic coherent state for a $2d$ droplet is simply the tensor product of states for the upper and lower Fermi surfaces and is given by
\begin{equation}
\begin{aligned}
    \ket{\t^a} \equiv \ket{\t^a_+, \t^a_{-}}&=\sum_{R_+,\bar R_{-}}s_{R_+}(\t^a_+)s_{\bar R_{-}}(\t^a_{-})\ket{R_+}\otimes \ket{\bar R_{-}}\\
    &=\sum_{S_+, S_{-}}\left( \sum_{R_+,R_{-}} \left(\sum_{S'}N^{R_+}_{S_+ S'}N^{R_-}_{S_{-}S'}\right)s_{R_+}(\t^a_+)s_{\bar R_{-}}(\t^a_{-}) \right)\ket{S_{+}\bar{S}_{-}}\\
    &=\sum_{S_+,S_-}\left(\sum_{S'}s_{S'}(\t^a_+)s_{S'}(\t^a_-) \right)s_{S_+}(\t^a_+)s_{S_{-}}(\t^a_{-})\ket{S_{+}\bar{S}_{-}}.
\end{aligned}
\end{equation}
Here we have used the fact that $s_{\bar R_-}(\t_-^a) = s_{R_-}(\t_-^a)$ up to a sign. Using the definition of coherent state one can show that the sum over $S'$ is given by
\begin{equation}
    \sum_{S'}s_{S'}(\t^a_+)s_{S'}(\t^a_-)= ({\t^a_+}|{\t^b_-}) = \exp\lb \sum_{n>0} \frac{\t_{+n}^a \t_{-n}^a}{n \hbar^2}\rb.
\end{equation}
This eventually gives us 
\begin{equation}\label{eq:fulltau}
    \ket{\t^a} \equiv \ket{\t^a_+, \t^a_{-}}=({\t^a_+}|{\t^b_-}) \sum_{S_+,S_-}s_{S_{+}}(\t^a_+)s_{S_{-}}(\t^a_{-})\ket{S_{+}\bar{S}_{-}}.
\end{equation}
Now, for a composite representation $\cS \equiv S_+\bar{S}_-$, we define a composite Schur function $s_{\cS}(\tau^a_+,\tau^a_-)$ as
\ben
s_{\cS}(\tau^a)\equiv s_{S_+\bar S_-}(\tau^a_+,\tau^a_-) = ({\t^a_+}|{\t^b_-}) s_{S_{+}}(\tau^a_+)s_{S_{-}}(\t^a_{-})
\een
and hence $\ket{\tau^a}$ in (\ref{eq:fulltau}) can be written as
\ben
\ket{\tau^a} =\sum_{\cS} s_{\cS}(\tau^a) \ket{\cS}.
\een
Using (\ref{eq:composite-basischange-inverse}) one can show that
\ben
e^{-\frac{1}{\hbar}H T} \ket{\cS} = e^{-\frac{\hbar}{2}C_2(\cS)T}\ket{\cS}
\een
Finally, evolution of $\ket{\tau^a}$ state to $\ket{\tau^b}$ state in time $T$ is given by,
\ben
\bra{\tau^b} e^{-\frac{1}{\hbar}H}\ket{\tau^a}  = \sum_{\cS} s_\cS(\tau^a) s_\cS(\tau^b) e^{-\frac{\hbar}{2}C_2(\cS)T}.
\een
This is the partition function of $2d$ Yang-Mills theory on cylinder specified by two holonomies corresponding to $\t^a$ and $\t^b$. As a special case one can check that when both $\t_+^a=\t_-^a=\t^*$ then we have (up to an overall constant)
\ben
s_{\cS}(\t^a) = \text{dim}S_+ \text{dim}\bar S_- = \text{dim}\cS
\een
and we get back the \emph{disc} amplitude. When all the $\ket \tau$ states are special \emph{i.e.} $\t_+^a=\t_-^a=\t_+^b=\t_-^b=\t^*$ we get back the sphere partition function of $2d$ Yang-Mills theory.

\section{Discussion}

In this paper we show the equivalence between unitary matrix quantum mechanics and $2d$ Yang-Mills theory and its variants via quantisation of phase space droplet. We explicitly construct the partition function of $2d$ Yang-Mills theory on a generic Riemann surface with gauge group $U(N)$ or \emph{q}-deformed $U(N)$ from the evolution of free Fermi droplets in one dimensional unitary matrix models. We note that the \emph{q}-deformation in the Yang-Mills side is related to the evolution of a particular types of geometries of droplets without deforming the gauge group associated with the matrix model. In that sense the droplet picture is more universal. The Hamiltonian as appearing in \eqref{eq:Hamab} also appears in earlier literature \cite{Natsuume:1994sp,Jevicki:1996fd}. However, they appear in the context of $2d$ bosonic string theory where the interesting cubic term can be interpreted as the splitting-joining interaction. In the current scenario we have arrived at the same by quantising the fluctuations over a \emph{droplet in two-dimensional phase space} that captures the eigenvalue distribution of $(0+1)$ dimensional unitary matrix model described by the action \eqref{eq:mmqmech}. Thus, our procedure gives a \emph{bottom-up} approach of arriving at \eqref{eq:Hamab}. In our work we have considered the potential $W(\theta)$ to be constant. It would be interesting to derive the phase space Hamiltonian for a generic potential and understand its meaning in the context of $2d$ Yang-Mills and string theory. Starting with a sufficiently generic action, quantisation of these fluctuations follows a Kac-Moody algebra. Our methodology, however has one restriction. The classical configuration of the Fermi surface that we start with has a quadratic profile. The small fluctuations introduced on this Fermi sea are ``small and shallow enough" so as not to destroy the quadratic profile. This ensures that a $\theta =$ constant line intersects the Fermi surface exactly twice validating \eqref{eq:BFdic}. Of course initial states with non-quadratic profiles are interesting in their own right but we will postpone such discussion to future works. It must however be noted that folds will generically form under time evolution even if we start with an unfolded Fermi surface where the collective field theory describing those non-interacting fermions is non-relativistic \cite{Das:1995gd,Alexandrov:2003ut,Das:2004rx}. Usually the fold states have non-zero quantum dispersion. Given a starting profile for the Fermi surface (say at $t=0$) \cite{Das:1995gd} explicitly calculated such fold formation times ($t_f$). We consider the evolution of coherent states \eqref{eq:coherentstate} constructed in section \ref{sec:phasestate} such that the evolution keeps the quadraticity of the Fermi surface at all times.

\acknowledgments

We would like to thank Suresh Govindarajan for fruitful discussions. We also acknowledged the illuminating discussion with Nabamita Banerjee, Suhas Gangadharaiah, Arnab Rudra, Ashoke Sen. The work of SD is supported by the grant no. \emph{EMR/2016/006294} and \emph{MTR/2019/000390} from the SERB, Government of India. SD also acknowledges the Simons Associateship of the Abdus Salam ICTP, Trieste, Italy. We also thank all the medical and non-medical workers who are working tirelessly in these troubled times. Finally, we are grateful to people of India for their unconditional support towards researches in basic sciences.

\appendix
\section{Eigenstates of phase space Hamiltonian}\label{app:eigenphase}

We will demonstrate explicitly that the state $\ket {R_+}$ as defined in \eqref{eq:YDbasis} is an exact eigenstate of the Hamiltonian \eqref{eq:Hamab}. \eqref{eq:Hamab} consists of two families of decoupled modes $a_n$ and $b_n$. For the sake of brevity we will only look at the action of the $a_n$ modes on $\ket {R_+}$. It is easy to see from \eqref{eq:KMprimary} that, the action of the part of the Hamiltonian denoted as $H_0$ in \eqref{eq:Hamat} on the Kac-Moody primary $\ket s$ is given by
\ben
H_0 \ket{s}=\left(-\frac{\hbar^2}{24}s+\frac{\hbar}{4}s^2+\frac{\hbar^2}{6}s^3\right)\ket{s}
\een
Using \eqref{eq:U1KM} and noting $a_{-n}=a_n^{\dagger}$, one can easily show $[a_n,(a_m^{\dagger})^{p}]=pn(a_n^{\dagger})^{p-1}\delta_{n,m}$. This further gives us the identity
\begin{equation}
     a_{-n}a_n\ket{\vec{k},s}=a_n^{\dagger}a_n \ket{\vec{k},s}=n\,k_n\ket{\vec{k},s}\ .
\end{equation}
Thus, the identity above along with \eqref{eq:YDbasis} leads to
\begin{equation}
\begin{aligned}
    H_{\text{free}}\ket{R_+}&=\frac{\hbar}{2}(1+2\hbar a_0)\sum_{n>0}a_n^{\dagger}a_n \ket{R_+}\\
    &=\frac{\hbar}{2}(1+2\hbar s)\sum_{n>0}nk_n\ket{R_+}
\end{aligned}
\end{equation}
where $l(R)=\sum_{n>0}nk_n$ is the total number of boxes associated to the Young diagram corresponding to the representation $R$. Thus, the "mostly zero vector" $\vec{k}$ can be interpreted as the cycle numbers of the Young diagram. 
The cubic part of the Hamiltonian \eqref{eq:Hamat} denoted by $H_{\text{int}}$, has a significantly more complicated action on $\ket {R_+}$. Action of $H_{\text{int}}$ on $R_+$ is given by
\begin{equation}
\begin{split}
    H_{\text{int}}\ket {R_+}& =\frac{\hbar^2}{2}\left[\sum_{m,n>0}(a_ma_na_{m+n}^{\dagger}+a_m^{\dagger}a_n^{\dagger}a_{m+n})\right] \ket{R_+}\\
    & =\frac{\hbar^2}{2}\left(l(R)+\sum_{i=1}^N(l_i^2-2il_i)\right) \ket {R_+}
\end{split}
\end{equation}
where $l_i$ happens to be number of boxes in each row of the corresponding Young diagram of the representation $R$.

\section{Reviewing $u(N)$ representations}\label{app:repgyan}
Any representation of the $u(N)=\left[su(N)\times u(1)\right]/\mathbb{Z}_N$ algebra can be represented by a pair $(R,Q)$ comprising of a standard Young diagram $R$ and a number $Q$, coming from the $u(1)$ generator of the algebra. However the $u(N)$ representations can also be written in terms of composite Young diagrams, specially in the $N\rightarrow\infty$ limit, which we discuss in the following sections.

\subsection{Composite representations}
As shown in \cite{Gross:1993hu}, a naive grouping of Young diagrams in terms of their total number of boxes under-counts the all possible $u(\infty)$ representations exactly by a factor of half. One can easily circumvent this problem by taking recourse to the notion of \emph{composite representations}.

\begin{figure}[H]
	\begin{center}
		\includegraphics[width=10cm,height=4cm]{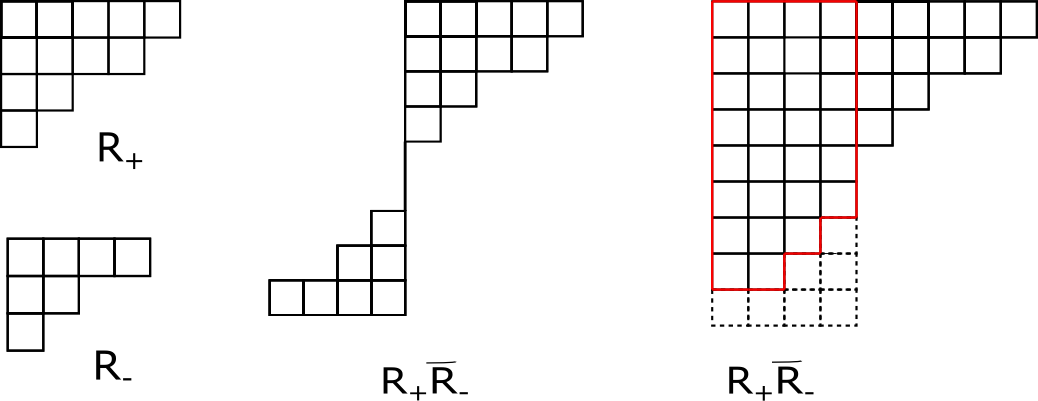}
		\caption{\footnotesize{An example of a composite representation for $N=9$. The red coloured border in the rightmost figure represents the boundary of the conjugate Young diagram corresponding to representation $R_-$.}}
		\label{fig:comp_rep}
	\end{center}
\end{figure}
Given two representations $R_+$ and $R_-$, a composite representation $R_+ \bar R_-$ is given by a Young diagram with boxes corresponding to $R_+$ placed on top right in a standard way and the boxes corresponding to $R_-$ placed upside down in the bottom left as \emph{anti-boxes}. The total number of rows in a Young diagram corresponding to a composite representation is always $N$ such that if the number of rows $r_+$ in $R_+$ and $r_-$ in $R_-$ do not add up to $N$, then $N-(r_+ + r_-)$ number of rows have zero length, in between $R_+$ and $R_-$ as shown in the centre diagram of Fig. \ref{fig:comp_rep}. Note that this procedure makes sense only in the large $N$ limit where none of the representations $R_+$ and $R_-$ has more than $N/2$ number of rows. In the $N\rightarrow\infty$ limit, summing over all possible representations of $u(N)$ is indeed equivalent to summing over all possible composite Young diagrams. 

This way of representing a composite diagram was used to demonstrate the factorization of $u(N)$ Yang-Mills theory into a chiral and anti-chiral sector in \cite{Aganagic:2005dh}. They further showed that this is equivalent to the composite representations of \cite{Gross:1993hu} where the conjugate Young diagram of $R_-$ is drawn first and then the diagram of $R_+$ is attached to it, on the right as depicted in the rightmost diagram of Fig. \ref{fig:comp_rep}.

\subsection{Quadratic Casimirs of $u(N)$ and $su(N)$}
For any semisimple Lie algebra $G$ with generators denoted by $\cT^a_\cR$ the quadratic Casimir $C_2(\cR)$ can be read off from the formula \cite{yellowbook,fulton-harris}
\begin{equation}\label{Qcasimirdef}
    C_2(\cR) \mathbb{1}_{\text{dim}\cR}=\sum_{a,b} [\cK\left(\cT^a_{\tiny\yng(1)},\cT^b_{\tiny\yng(1)}\right)]^{-1} \cT^a_\cR\cT^b_R,
\end{equation}
with $\cK\left(\cT^a_{\tiny\yng(1)},\cT^b_{\tiny\yng(1)}\right)$ being the Killing form where ``$\tiny\yng(1)$" symbolizes the fundamental representation. Interestingly one can also check that quadratic Casimir is the same for a representation and its conjugate. For the case of $G\equiv su(N)$, with $a\in [1,N^2-1]$, one can simplify the quadratic Casimir by characterising the irreducible representations $R$ of $su(N)$ by a standard Young diagram with row lengths $l_i$ where $\left(i\in[1,N-1]\right)$ with the condition $\infty\geq l_1\geq l_2\geq\cdots\geq l_{N-1}\geq 0$. In that case, \eqref{Qcasimirdef} simply reads
\begin{equation}
    C_2(R)=N l(R)+\kappa_R- {l(R)^2\over N};\quad l(R)=\sum_{i=1}^{N-1}l_i;\quad \kappa_R=l(R)+\sum_{i=1}^{N-1}(l_i^2-2il_i).
\end{equation}
 Since $u(N)=\left[su(N)\times u(1)\right]/\mathbb{Z}_N$, the eigenvalue $Q$ of the $u(1)$ generator, appropriately called the \emph{charge}, is constrained to be equal to $\big(l(R)\text{ mod } N\big)$. Again resorting to \eqref{Qcasimirdef}, the quadratic Casimir for $u(N)$ algebra can be written as 
\begin{equation}\label{uNCasimir}
    C_2(R,Q)=C_2(R)+{Q^2\over N}=N l(R)+\kappa_R+2s l(R)+Ns^2,
\end{equation}
where the second equivalence stems from the relation $Q=l(R)+Ns$ with $s\in \mathbb{Z}$ which is just a simple manifestation of the fact that $Q=l(R)\text{ mod }N$. We have chosen this particular basis of the $u(N)$ in this paper because of its apparently simple relation with the $su(N)$ irreducible representation $R$. Interestingly one can go a step further with the $u(N)$ representations $(R,Q)$ and introduce coupled representations \cite{Gross:1993hu,Naculich:2007nc} or the \emph{extended} Young diagram $\cR$ which may or may not have negative number of boxes (also termed as ``anti-boxes" \cite{Aganagic:2005dh}) as discussed in the previous section.\\
The origin of these \emph{exotic} Young diagrams relies on the clever rewriting of \eqref{uNCasimir} as
\begin{eqnarray}
    &&C_2(R,Q)=N \baR{l}(R)+\baR{\kappa_R}\nonumber\\
    &&\baR{l}(R)=\sum_{i=1}^{N}\bar{l_i};\quad\baR{\kappa_R}=\baR{l}(R)+\sum_{i=1}^{N}(\baR{l_i}^2-2i\baR{l_i})\\
    &&\baR{l_i}=l_i+s;\quad \baR{l_N}=s;\quad s\in\mathbb{Z}.\nonumber
\end{eqnarray}
Therefore one can define an extended Young diagram $\cR$ with boxes in the $i^{\text{th}}$ row being $\baR{l_i}$ with the condition that $\infty\geq \baR{l_1}\geq\baR{l_2}\geq\cdots\geq \baR{l_N}\geq-\infty$. Hence the charge is now simply the total number of boxes of $\cR$ as $Q=\sum_{i=1}^{N}\baR{l_i}$. Therefore the quadratic Casimir of $u(N)$ now reduces to the relation
\begin{eqnarray}
    C_2(\cR)=NQ+\baR{\kappa_{\cR}}; \quad \baR{\kappa_{\cR}}=Q+\sum_{i=1}^{N}(\baR{l_i}^2-2i\baR{l_i}).
\end{eqnarray}

\section{Twisted surgery and higher point functions}
\label{app:higherpoint}

We consider time evolution of a classical shape $\ket{\tau^a}$ to $\ket{\tau}$ in time $T_1$ - such evolution is given by a cylinder amplitude $\cC(\t^a, \t,T_1)$. We then consider another amplitude - transition from a classical shape $\ket{\tau}$ to $\ket{\tau^b}$ from $T_1$ time to $T_2$, $\cC(\t,\t^b,T_2-T_1)$. If we glue these two amplitudes along $\tau$ circle and integrate over all possible $\tau$ states we get an amplitude for transition from $\ket{\tau^a}$ shape to $\ket{\tau^b}$ shape in time $T_2$. However one can glue these two cylinders along $\t$ circle after giving a local twist. The twists is given in $\theta$ plane. A $\ket{\tau}$ state has a image in $\theta$ plane. We glue $\theta_i$ point of the first cylinder with $\theta_i$ point of the second cylinder after giving a local twist $\phi^c_i$ and then we integrate over all $\theta_i$ points i.e. $\ket{\tau}$ state. We call this $\vec \phi^c$ twisted amplitude $\cP(\t^a,\vec \phi^c,\tau^b)$. Thus we have
\ben
\cP(\t^a,\vec \phi,\tau^b)= \int [d\theta] W(\theta_i) \sum_{RR'} s_R(\tau^a) s_R(\theta_i+\phi^c_i)  s_R(\theta_i) s_R(\tau^b) e^{-\frac{\hbar}{2} (C_2(R)T_1+C_2(R')(T_2-T_1))}.
\een
We use the identity
\ben
\int [d\theta] W(\theta_i) s_R(\theta_i+\phi^c_i) s_{R'}(\theta_i) = \frac{s_R(\phi^c_i)}{\text{dim}_R}\delta_{RR'}
\een
Note that in the large $N$ limit a local twist $\{\phi^c_i\}$ corresponds to a distribution $\sigma(\phi^c)$ and hence there exists a corresponding coherent state $\ket{\tau^c}$. Therefore we write $s_R(\vec \phi^c) \ra s_R(\tau^c)$ and denote the above amplitude by $\cP(\tau^a,\tau^c,\tau^b,T)$.  Thus we have
\ben
\cP(\tau^a,\tau^c,\tau^b,T)= \sum_{R} \frac{s_R(\tau^a)s_R(\tau^b)s_R(\t^c)}{\textrm{dim}R} e^{-\frac{\hbar}2 C_2(R)T_2}.
\een
As a consistency check, if we set the twist parameters $\phi^c_i=0$ i.e. $\ket{\t^c}=\ket{\t^*}$, we get back the cylinder amplitude as expected. Therefore from the point of view of evolution of droplet, one may think that an initial shape $\ket{\t^a}$ evolves to a final shape $\ket{\t^b}$ in time $T_2$ with an external twist $\vec \phi^c$ at some intermediate time $0<T_1<T_2$. 

\section{Villain action}\label{app:villain}

The abelian $U(1)$ Villain action is just a Jacobi theta function
\ben
\text{exp}(-S_{V}(\theta))=\sum_{l=-\infty}^{\infty}e^{-\frac{1}{g^{2}}(\theta + 2\pi l)^{2}};\quad l\in \mathbb{Z}.
\een
It was first introduced in \cite{Villain:1975} as an approximation to the Hamiltonian of a $2d$ planar classical magnet and is well-known to be used in the study of planar Heisenberg model \cite{Susskind_ph:1979}. Owing to the analogy between $2d$ planar model and four-dimensional abelian gauge theory \cite{Kogut_review}, it also appears in the study of lattice gauge theories and therefore its appearance in the context of $2d$ Yang-Mills theory is not that much of a surprise. Generalization of Villain's action with an $U(N)$ gauge group is given by
\ben\label{eq:Villaine}
\text{exp}(-S_{V}(\{\theta_{i}\})=\prod_{i=1}^{N}\sum_{{l_{i}}=-\infty}^{\infty}\text{exp}[-\frac{N}{\lambda}(\theta_{i}+2\pi l_{i})^{2}]
\een
where $\theta_{i}$'s are the invariant angles of $U\in U(N)$. As the form of (\ref{eq:Villaine}) suggests, one can expand this action in the $U(N)$ character basis following \cite{Onofri:1981qk} as
\ben \label{eq:unvilchar}
\label{eq:ch_expansion}
\frac{\text{exp}[-S_{V}]}{Z}=\sum_{\cR}c_{\cR}\text{Tr}_{\cR}(U)
\een
where $\text{Tr}_{\cR}(U)$ is the unitary group characters and $Z$ is the two-dimensional lattice gauge theory partition function with generalized Villain's action. The coefficients $c_{\cR}$ can then be found using the orthogonality of characters. After a little rearrangement which can be written as
\ben
\label{eq:ch_coefficient}
c_{\cR}=q^{\sum_{j}(l_{j}-j+1)^2}\prod_{j>k}\left[\frac{1-q^{2(l_{j}-l_{k})}}{1-q^{2(j-k)}}\right] \qquad (q=e^{-\lambda/4N})
\een
where the integers $ {l_{i}} $ are related to the number of boxes in a Young diagram corresponding to the representation $\cR$ of $U(N)$. In \cite{Romo_2012} it was shown that the character expansion (\ref{eq:ch_expansion}) with coefficients given by (\ref{eq:ch_coefficient}) reduces to the $q$-deformed disc amplitude of 2D Yang-Mills theory (\ref{eq:qddisc}). Generalization of Villain's action to $U(N)$ also leads to the heat kernel (\ref{eq:heatkernel}) as pointed out in \cite{Menotti:1981ry}.

Another interesting aspect of writing the character expansion of Villain action is the gluing of two Yang-Mills theories with different $q$-deformations. If one considers generalized Villain's action with 't Hooft couplings $\lambda_{1}$ and $\lambda_{2}$ then the integral
\ben
\int dU \frac{\text{exp}[-S_{V}(U,\lambda_{1})]}{Z(\lambda_{1})}\frac{\text{exp}[-S_{V}(U,\lambda_{2})]}{Z(\lambda_{2})} = \sum_{R}c_{R}(q_{1})c_{R}(q_{2})
\een
is equivalent to gluing two disc amplitudes with deformations $q_{1}$ and $q_{2}$ respectively. Using (\ref{eq:ch_coefficient}) and following \cite{Romo_2012}, the above expression can be reduced to
\ben
\sum_{R} \text{dim}_{q_{1}}R \text{ dim}_{q_{2}}R \hspace{3pt} q_{1}^{\frac{1}{2} C_{2}(R)} q_{2}^{\frac{1}{2} C_{2}(R)}.
\een

\bibliography{PlancherelNotes.bib}

\end{document}